\documentclass[preprint,prb,showpacs,preprintnumbers,amsmath,amssymb]{revtex4}

\usepackage{graphicx}% Include figure files
\usepackage{dcolumn}% Align table columns on decimal point
\usepackage{bm}% bold math
\usepackage{color}

\begin{document}

\title{Photo-gating Carbon Nanotube Transistors}

\author{Matthew S. Marcus}
 \affiliation{Department of Physics, University of Wisconsin - Madison}
\author{J. M. Simmons}
 \affiliation{Department of Physics, University of
 Wisconsin - Madison}
\author{O. M. Castellini}
 \affiliation{Department of Physics, University of
 Wisconsin - Madison}
\author{R. J. Hamers}
 \affiliation{Department of Chemistry, University of Wisconsin -
Madison}
\author{M. A. Eriksson}
 \affiliation{Department of Physics, University of Wisconsin -
 Madison }

\begin{abstract}
 Optoelectronic measurements of carbon nanotube transistors have
shown a wide variety of sensitivites to the incident light.  Direct
photocurrent processes compete with a number of extrinsic
mechanisms.  Here we show that visible light absorption in the
silicon substrate generates a photovoltage that can electrically
gate the nanotube device.  The photocurrent induced by the changing
gate voltage can be significantly larger than that due to direct
electron-hole pair generation in the nanotube.  The dominance of
photogating in these devices is confirmed by the power and position
dependence of the resulting photocurrent.  The power dependence is
strongly non-linear and photocurrents are measured through the
device even when the laser illuminates up to 1~mm from the nanotube.
\end{abstract}


\maketitle

\section{\label{sec:level1}Introduction}
Since their discovery, carbon nanotubes have shown great promise for
a wide variety of applications including nanoelectromechanical
systems\cite{McEuenNEMS,ZettlNEMS} and chemical and biological
sensing.\cite{LeeNanoLett,StarPNAS}  The observation of
photoconductivity in carbon nanotubes has opened a number of
avenues of research in both characterization and photonic
applications of
nanotubes.\cite{FujiwaraJJAP,LevitskyAPL,FreitagNanoLett}
Photocurrents have been used to characterize the electronic
structure of nanotubes, such as the energies of van Hove
singularities,\cite{FreitagNanoLett,OhnoAPL,MohiteCPL} as well as to
image the Schottky barriers between a semiconducting nanotube and
the metal contacts,\cite{BalasubAPL} and to build novel
photodetectors using suspended nanotubes.\cite{HaoChihProc}
Recently, theory and experiments have shown that direct
photogeneration of carriers is unable to explain the full
photophysics of carbon nanotubes, and that
excitonic\cite{MeleSSC,MaJPCB,ShengPRB} and environmental (both
substrate and gaseous)\cite{ItkisScience} interactions must also be
considered. An interesting feature of illuminating nanotube devices
on a silicon/silicon dioxide substrate is that the silicon can also
absorb light, leading to a voltage at the Si/SiO$_2$ interface. Such
photo-voltages can, in many cases, be the dominant component of the
induced photocurrent in nanotube devices. Understanding the
interaction between the incident light, the back-gate, and the
nanotube (photo)transistor is important both for potential
applications of such devices, and to enable fundamental
understanding of the direct nanotube-light interaction in nanotube
transistors.

We show that photo-voltages in the silicon gate of carbon nanotube
transistors have a significant impact on the measured photocurrent.
 The measured photocurrent is independent of laser polarization,
indicating that direct electron-hole pair generation in the nanotube
is negligible. We show that modulating the incident light
effectively adds a small ac gate voltage, revealing the derivative
of the source-drain current $I_{ds}$ in the photocurrent. Comparing
the numerical derivative of the $I_{ds}$ vs.\ gate voltage curve
taken without the light to the measured photocurrent as a function
of gate voltage, we are able to extract the induced photovoltage
which is typically on the order of $-10$ to $-100$~mV. Consistent
with the photogating effect, the power and position dependence of
the measured photocurrent is extremely non-linear. Indeed,
photocurrents persist even when the laser illuminates up to 1~mm
away from the nanotube position.  The experiments are consistent
with a theoretical analysis of the photo-gating mechanism.  These
measurements suggest that photo-voltages produced in transistor
gates may provide the basis for unusual types of photo-detectors:
devices in which the light absorbed by the gate produces a
modulation in a transistor response, as opposed to the direct
collection of photo-generated carriers in more conventional
photo-detectors.

\section{\label{sec:level2}Experimental}
The carbon nanotubes in the devices studied here are grown by
chemical vapor deposition (CVD).  Islands of nanotube catalyst are
defined using electron-beam lithography,\cite{KongNature} and the
nanotubes are grown at 900~$^\text{o}$C using a feedstock of methane
(400~sccm) with a co-flow of hydrogen
(20~sccm).\cite{MelechkoJAP,MoisalaJPCondMat,FirstNano,SimmonsSmall}
The ends of the nanotube are buried under Ti/Au using electron-beam
lithography in order to define the source-drain contacts.  The
nanotube channel and source-drain leads rest on top of a 500~nm
thick SiO$_2$ layer that acts as a gate dielectric where the back
gate is a heavily doped (nominally $\sim 10^{18}$~cm$^{-3}$) p-Si
substrate. Nanotube transistors have been fabricated with both short
($\sim5~\mu$m) and long ($\sim500~\mu$m) channels.
\begin{figure}
 \centering
 \includegraphics[width=3.25in]{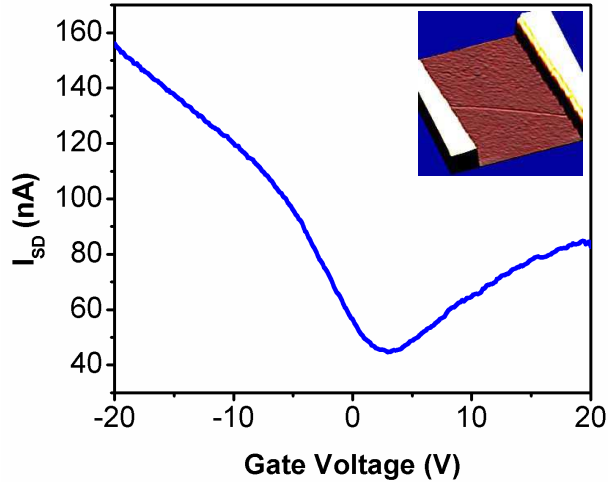}
 \caption{Source-drain ($I_{SD}$) as a function of back gate voltage.  The conduction through the nanotube increases at
 large values of both negative and positive gate voltages.  Inset:  AFM image of a $\sim5~\mu$m long single wall carbon nanotube
 transistor device.}
 \label{IVg}
\end{figure}
Fig. \ref{IVg} is a plot of the source-drain current ($I_{SD}$) in a
short channel device as a function of gate voltage under a fixed
source-drain voltage ($V_{SD}=250$~mV).  The conductance of the
nanotube shown is dependent on the gate voltage, and it shows
increased conductance at both positive and negative gate voltages.
The gate dependence could indicate the presence of a semiconducting
nanotube or may indicate effects of two nanotubes in parallel, small
metallic gap openings,\cite{OuyangScience} or resonant electron
scattering.\cite{BockrathScience,BozovicPRB}  Regardless of the
origin of the gate dependence, all nanotubes that have a gate
dependence can respond to photovoltages induced in the gate.

Nanotube devices are illuminated with a $P_o=4$~mW HeNe laser that
has a photon energy ($E_{\gamma}=1.96$~eV) which is larger than both
the band gap of our nanotubes\cite{BandGaps} as well as the band gap
of the Si gate (E$_{G,Si}=1.12$~eV).  For short channel devices the
laser always illuminates the entire nanotube device, including the
bulk of the nanotube and the metal-nanotube interface. The laser is
modulated using an optical chopper and all photocurrents are
measured with a lock-in amplifier referenced to the chopping
frequency.  When the signal is modulated with a chopper, the
photo-voltage is a step function at the on or off transition,
producing a capacitive spike in the current.  Once the photo-voltage
has reached a steady state and is not changing with respect to time,
the capacitive current is zero.  Slow modulation frequencies
(10-100~Hz) are used in order to minimize the effects of capacitive
coupling between the source-drain contacts and the gate during the
turn on and turn off transitions.\cite{FreitagNanoLett}

\begin{figure}
 \centering
 \includegraphics[width=3.25in]{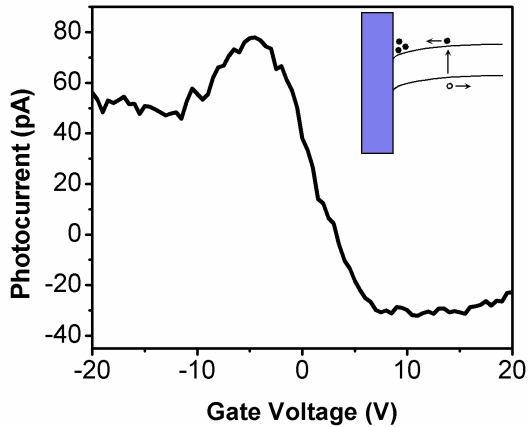}
 \caption{Photocurrent measured through the nanotube channel as a function of gate voltage.
  The photocurrent switches sign near $V_g=0$.
  Inset:  Schematic of the energy bands with initial band bending at the Si/SiO$_2$ interface.}
 \label{PCVg}
\end{figure}
Photocurrents may arise from several sources.  These include direct
processes, such as electron-hole pairs generated in the nanotube, or
indirect processes such as photo-desorption\cite{ChenAPL,ShimAPL} or
photo-gating.  The photocurrent for a nanotube transistor as a
function of gate voltage is shown in Fig.~\ref{PCVg} and,
intriguingly, changes sign as a function of gate voltage. Also, the
photocurrent shows no dependence on the polarization of the laser
with respect to the nanotube axis.\cite{FreitagNanoLett,BalasubAPL}
The change in sign of the photocurrent in Fig.~\ref{PCVg} and the
absence of polarization dependence indicates that direct processes
are not the source of the observed photocurrent. Direct photocurrent
in a biased nanotube would have the same sign for all applied gate
voltages because the electric field provided by the bias voltage
would sweep all the photo-generated carriers in the same direction,
independent of the gate voltage. Photo-desorption of oxygen from the
metal-nanotube contacts also cannot account for the observed
photocurrent.  The removal of oxygen from the nanotube and/or
contacts, thereby affecting the metal-nanotube Schottky barrier,
would lead to decreased p-channel conductivity (negative gate
voltages), and increased n-channel conductivity (positive gate
voltages).\cite{ShimAPL} The data in Fig.~\ref{PCVg} show an
increase in p-channel conductivity, and a decrease in the n-channel
conductivity, the opposite of what is expected for a
photo-desorption process.  Photo-gating, as we show below, can
explain both the increased p-, and decreased n-channel conductivity
as well as the observed sign change.

\section{\label{sec:level3}Photo-gating Mechanism}
Photo-gating occurs when photo-excited charge in the silicon gate
causes a change in potential ($\delta V_{PV}$) at the Si/SiO$_2$
interface.  In order for photo-gating to occur, there needs to be a
slope in the Si energy bands at the Si/SiO$_2$ interface. Bending in
the Si energy bands provides an electric field that separates the
electron-hole pairs produced by the laser and a potential well to
trap either electrons or
holes.\cite{SchroderMeasSciTech,KronikSurfSciRep}  Band bending is
often present due to charge in the oxide (mobile ions, fixed charge,
and interface states) which induces a depletion region near the
Si/SiO$_2$ interface.  In addition to the presence of band bending,
to produce a significant photovoltage the potential well must be
deep enough to act as a trap at the measurement temperature
($\sim25$~meV).

In order to determine if there is band bending at the Si/SiO$_2$
interface, capacitance-voltage (CV) characteristics of a
metal-oxide-semiconductor capacitor (MOSC) are
measured.\cite{SchroderBook,SzeBook} The MOSC is fabricated on an
oxidized Si wafer from the same batch as the nanotube samples
studied.  Fits to the experimental CV data (not shown) indicate that
the bulk doping concentration is
$N_a\approx1.2\times10^{18}~\text{cm}^{-3}$, and that there is an
offset voltage $\Delta V_o=-28$~V.  The non-zero offset voltage
indicates the presence of initial band bending due to oxide charge.
If the offset voltage were zero there would be no curvature in the
Si energy bands and therefore no photo-gating effect.  The oxide
charge density can be estimated using $N_{ox}\approx -C_{ox}\Delta
V_o/e\approx 1.2\times10^{12}~\text{cm}^{-2}$ where C$_{ox}$ is the
capacitance per area of the oxide, and $e$ is the magnitude of the
electron charge.  The positive oxide charge induces an equal
negative charge in the silicon substrate ($Q_s=
-N_{ox}\approx-1.2\times10^{12}~\text{cm}^{-2}$), such that the
energy bands in the silicon are bent downwards, establishing a trap
for photo-excited electrons.

\begin{figure}
 \centering
 \includegraphics[width=3.25in]{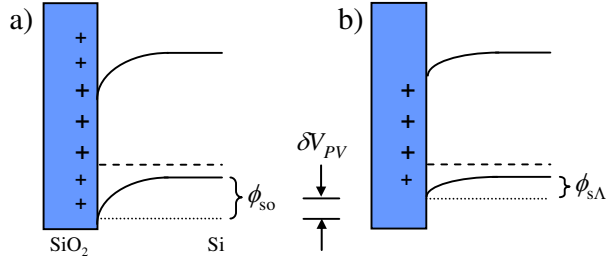}
 \caption{Schematic band diagrams showing band bending $\phi_s$ in the silicon substrate for the as fabricated device (a) and under
 laser illumination (b).  For typical oxide charge densities, the initial band bending $\phi_{so}$ is $\sim100-200$~mV.
 Under illumination, the positive oxide charge is partially screened by photogenerated
 electrons that become trapped at the interface, reducing the surface potential from
 $\phi_{so}\rightarrow\phi_{s\Lambda}$.  The resulting photovoltage $\delta V_{PV}$ is given by Eq.~\ref{eqnVPV}.}
 \label{BandDiag}
\end{figure}

\begin{figure}
 \centering
 \includegraphics[width=3.25in]{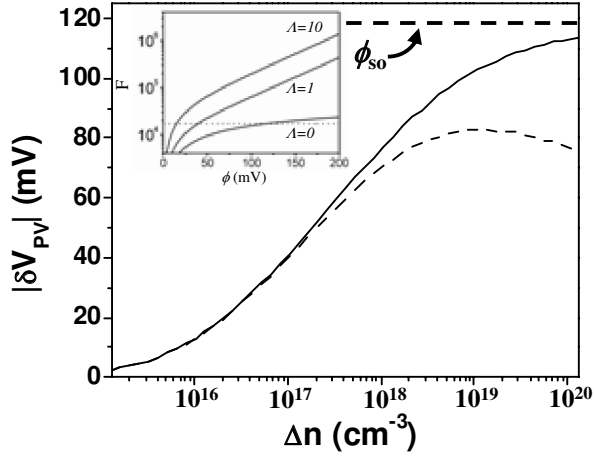}
 \caption{Calculated magnitude of photovoltage as a function of the
  number of photo-excited carriers, assuming a bulk doping density of
  $N_a\approx1.2\times10^{18}~\text{cm}^{-3}$ and an initial surface
  potential of $\phi_{so}\sim120$~mV.  The solid line includes only
  the effect of the initial band bending.  The dotted line includes
  the effect of the Dember voltage (see appendix).
  Inset:  Calculated normalized electric field at the Si/SiO$_2$ interface as a function of the surface potential and injection ratio.
  The dotted line indicates the value of the field $F_o$ as determined from the initial charge in the silicon $Q_s$.}
 \label{PVdN}
\end{figure}

In order to produce a photovoltage, the magnitude of the potential
created as a result of oxide charge must be significant enough to
provide an electric field that can redistribute the electron-hole
pairs generated by the laser. The charge in the silicon is a
function of the surface potential (band bending, see Fig.~\ref{BandDiag}a) $\phi_s$ and is given by the
expression\cite{SchroderMeasSciTech}
\begin{equation}
  Q_s=-\sqrt{2k_BT \varepsilon_{Si} n_i}F(U_s,U_b,\Delta n)
  \label{eqnQs}
\end{equation}
where
\begin{eqnarray}
 F(U_s,U_b,\Delta n)=&&[e^{U_b}(e^{-U_s}+U_s-1)\nonumber\\
 &&+e^{-U_b}(e^{U_s}-U_s-1)\nonumber\\&&+\Lambda
 e^{U_b}(e^{U_s}+e^{-U_s}-2]^{1/2} \label{eqnF2}
\end{eqnarray}
is the normalized interfacial electric field, assuming that the
depletion width is smaller than the diffusion length, and that
recombination currents can be ignored.  In the previous equations
$k_B$ is Boltzmann's constant, $T$ is temperature, $\varepsilon
_{Si}$ is the dielectric constant of Si, $n_i$ is the intrinsic
number of carriers, $U_s$ and $U_b$ are the normalized surface and
bulk potential $U_s=q\phi_s/k_BT$, $U_b=\ln(N_a/n_i)$ and $\Lambda$
is the ratio of photo-excited carriers to the equilibrium number of
majority carriers $\Lambda=\Delta p/p_o=\Delta n/p_o$, also referred
to as the injection ratio. Due to the complexity of Eq.~\ref{eqnF2},
the magnitude of the potential is most easily found graphically or
numerically. First, in the absence of light, $F(U_s,U_b,\Delta n)$
becomes a function only of the surface potential $\phi_s$.  Using
the value of $Q_s$ determined above, $F(U_s,U_b,\Delta n=0)\equiv
F_o$ is computed using Eq.~\ref{eqnQs}. The value of $F_o$ is shown
as the dotted line in the inset of Fig.~\ref{PVdN}.
Next, using Eq.~\ref{eqnF2}, $F$ is plotted as a function of
$\phi_s$ and the intersection of the curve with the dotted line sets
the initial amount of potential $\phi_{so}$ due to the oxide charge.
For the MOSC, the potential $\phi_{so}$, shown schematically in Fig.~\ref{BandDiag}a, is measured to be
$\sim120$~mV which is larger than the thermal carrier energies,
enabling photo-generated electrons to be trapped at the Si/SiO$_2$
interface.

Photo-excited electrons and holes in the silicon are separated by
the interfacial electric field. This charge redistribution changes
the potential in the silicon from $\phi_{so}$ to a different value
$\phi_{s\Lambda}$ (Fig.~\ref{BandDiag}b). The difference between the two potentials is the
measured photo-voltage,
\begin{equation}
 \delta V_{PV}=\phi_{s\Lambda}-\phi_{so}.
 \label{eqnVPV}
\end{equation}
So long as the laser does not affect the amount of oxide charge, the
total charge in the silicon remains constant and, according to
Eq.~\ref{eqnQs}, the value of $F$ must stay fixed at $F_o$.  When
illumination causes $\Lambda$ to become nonzero, $U_s$, and
therefore $\phi_s$, must change to keep $F=F_o$.  The inset of
Fig.~\ref{PVdN} shows several different $F$ vs. $\phi_s$ curves as
the injection ratio $\Lambda$ is varied. The potential under
illumination $\phi_{s\Lambda}$ is the intersection point between the
curve for a given $\Lambda$ and the initial interfacial field $F_o$.
For low values of the injection ratio, there are few photo-generated
carriers to fill the potential well at the Si/SiO$_2$ interface, and
$\phi_{s\Lambda}\approx\phi_{so}$. As the injection ratio is
increased, there are more carriers to fill the well, and the bands
become less curved, reaching flat-band conditions for high values of
$\Lambda$. Thus, as the injection ratio increases, the measured
photovoltage will change from 0 to $-\phi_{so}$.
The magnitude of the resulting photo-voltage is plotted as a function of
photo-carrier density in Fig.~\ref{PVdN}, using a bulk silicon
doping density of $N_a\sim 1.2\times10^{18}~\text{cm}^{-3}$ ($\Delta
n=\Lambda N_a$) and the measured initial surface potential
$\phi_{so}\sim120$~mV.  Once the bands are flat
($\phi_{s\Lambda}=0$), there is no electric field to separate the
electrons and holes and the measured photovoltage saturates.  This is illustrated in Fig.~\ref{PVdN} where the
photovoltage asymptotically approaches $\phi_{so}$ when $\Lambda$
becomes large, and in the inset where the interface potential tends
to 0. It is important to note that since different processing
conditions can lead to changes in the initial surface potential,
photovoltages measured on different samples can have higher or lower
values than the maximum presented in
Fig.~\ref{PVdN}.\cite{SchroderBook} Also, for sufficiently high
injection ratios ($\Delta n\gtrsim p_o$), other processes, such as
the Dember voltage\cite{KronikSurfSciRep} presented in the appendix,
can become important and could dominate over the photovoltage due to
initial band curvature as shown by the dashed curve in
Fig.~\ref{PVdN}. As we show next, we remain in the low-injection
limit where these additional effects can be ignored.

In the limit of low-injection conditions, $\Delta n<p_o$, we can
estimate the number of excited carriers as\cite{SchroderMeasSciTech}
\begin{equation}
 \Delta n\approx \frac{I\alpha L_n}{(1+\alpha
 L_n)(s_1+D_n/L_n)}\lesssim 10^{17}~\text{cm}^{-3},
 \label{eqnLowInjection}
\end{equation}
where $\alpha \approx 3000~\text{cm}^{-1}$ is the absorption
coefficient\cite{SzeBook} of Si at $\lambda~=~633$~nm,
$D_n=7~\text{cm}^2$/s is the minority carrier diffusion coefficient,
$L_n=45~\mu$m is the diffusion length, $I$ is the photon intensity,
and $s_1$ is the surface recombination velocity at the Si/SiO$_2$
interface. The diffusion coefficient ($D_n=k_BT/q\mu_n$) and
diffusion length ($L_n=\sqrt{D_n\tau_n}$) are calculated using the
minority carrier recombination lifetime,\cite{TyagiSolidStateElect}
$\tau_n = 3\text{x}10^{-6}$~s, and the minority carrier drift
mobility,\cite{SzeBook} $\mu_n\sim270~\text{cm}^2$/Vs. The photon
intensity is given by $I=\xi\eta(1-R)2P_o/eE_\gamma\pi\omega^2$,
where $\eta$ is the quantum efficiency (assumed to be unity),
$R=0.5$ is the measured reflectivity of the Si, and $\omega\sim
20\mu$m is the estimated laser spot radius. The factor of $\xi=0.2$
is used to account for the portion of the laser power incident on
the underlying silicon that is not absorbed by the metal leads. The
surface recombination velocity depends on a variety of factors
including interface state density, gate voltage, and the number of
photo-excited carriers,\cite{AberleJAP} and is typically on the
order of $100-1000$~cm/s. As an order of magnitude estimate, we
assume that that $s_1$ is negligible, which is valid as long as
$s_1<D_n/L_n\approx 1500$~cm/s. Due to our approximations for $s_1$
and $\eta$, and because Eq.~\ref{eqnLowInjection} does not account
for diffusion of carriers parallel to the Si/SiO$_2$ interface, the
calculated value is an over-estimate of the photo-excited carrier
density.  This over-estimate confirms that we are in the
low-injection limit ($\Delta n<N_a\approx10^{18}~\text{cm}^{-3}$)
for our experimental conditions. Using $\Delta n$ and
Fig.~\ref{PVdN}, the photo-voltage is estimated to be $\delta
V_{PV}<-40$~mV. The actual photo-voltage measured in the experiment
is extracted and is compared to this value later.

\section{Discussion}
The photo-voltage produced in the silicon gate couples into the
nanotube circuit by providing an additional gate voltage for the
nanotube channel.  When the light is modulated, the effective gate
voltage is $V_g=V_{DC}+\delta V_{PV}(t)$, where $\delta V_{PV}$ is
the time dependant photovoltage and $V_{DC}$ is an existing gate
voltage applied by a battery.  Since the current in the channel is
measured with a lock-in amplifier, the measured photocurrent is
$I_{PC}=\frac{\partial I_{SD}}{\partial V_g}\delta V_{PV}$. Thus,
the data shown in Fig.~\ref{PCVg} may be understood as a derivative
measurement of Fig.~\ref{IVg}, and the photocurrent reveals the
change in the source-drain current due to the addition of a small
negative ac gate voltage to the existing dc gate voltage. From
Fig.~\ref{IVg} we see that adding a small negative gate voltage to
an existing negative gate voltage causes the source-drain current to
increase, yielding a positive derivative and positive photocurrent.
At positive gate voltages, the effect of adding a small negative
gate voltage causes the source-drain current to decrease, yielding a
negative photocurrent.

\begin{figure}
 \centering
 \includegraphics[width=3.25in]{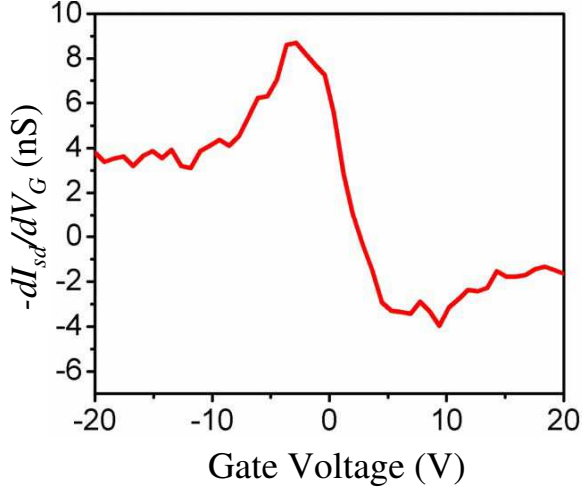}
 \caption{Numerically calculated derivative from the data in Fig.~\ref{IVg}.  The form of the derivative is extremely
 similar to the measured photocurrent in Fig.~\ref{PCVg} where the p-channel conduction is increased and the n-channel is suppressed.}
 \label{NumDerivative}
\end{figure}

The magnitude of the photo-gated photocurrent depends on both the
nanotube-gate coupling $\frac{\partial I_{SD}}{\partial V_g}$ and
the photo-voltage $\delta V_{PV}$ at the Si/SiO$_2$ interface. The
photo-voltage can be extracted by dividing the photocurrent signal
by the gate derivative $\frac{\partial I_{SD}}{\partial V_g}$, which
is calculated numerically from the data in Fig.~\ref{IVg} and is
shown in Fig.~\ref{NumDerivative}.
The photo-voltage for this device is $\delta V_{PV}=-15$~mV and is
independent of the applied gate voltage.  This measured value of
photovoltage is similar to the over-estimate of calculated above,
supporting the photogating mechanism.

\begin{figure}
 \centering
 \includegraphics[width=3.25in]{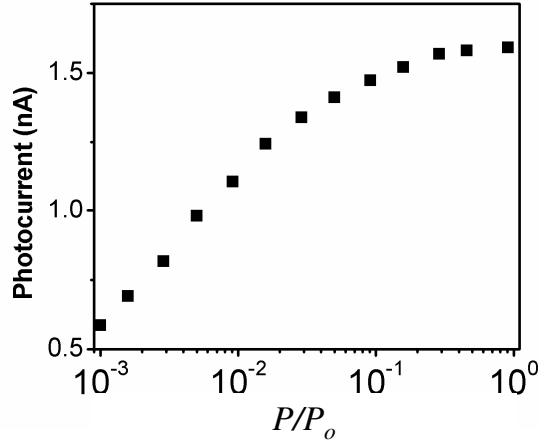}
 \caption{Measured photocurrent as a function of the incident laser power on a long channel nanotube device.
  The non-linearity of the intensity dependence indicates that the photocurrent does not arise from a direct
  process.}
 \label{PCvsIntensity}
\end{figure}

The power dependence of the photocurrent is another important
parameter for establishing the role of photo-gating.  The generation
rate of photo-carriers is linear with laser power, both in the
nanotube and in the silicon.  For a direct photocurrent process in
the nanotube $I_{PC}$ is expected to be proportional to the
generation rate, whereas for photo-gating $I_{PC}\propto\delta
V_{PV}$ which is non-linear in the number of excited carriers. Using
neutral density filters to change the incident laser intensity, the
photocurrent as a function of power was measured on a long channel
nanotube device\cite{IronNitrateGrowth} and is shown in
Fig.~\ref{PCvsIntensity}.  Using the nanotube-gate coupling and the
measured photocurrent, the induced photovoltage for this device was
measured to be $\delta V_{PV}=-160$~mV.
Similar to Fig.~\ref{PVdN}, the photocurrent changes logarithmically
under low injection before saturating at high laser intensities.

\begin{figure}[tph]
 \centering
 \includegraphics[width=3.25in]{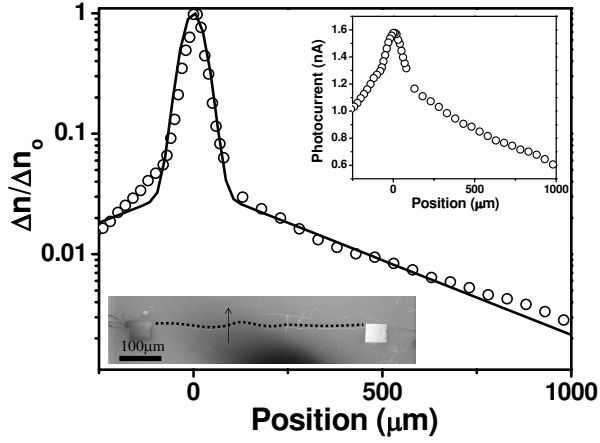}
 \caption{Ratio of photo-generated electrons in the silicon substrate as a function of position.  The circles are the
 experimentally measured values, extracted by relating the raw photocurrent vs. position in the upper inset to the intensity dependence
 shown in Fig.~\ref{PCvsIntensity}.  The line is a fit to Eq.~\ref{eqnGenerationMobility} which includes direct generation
 of carriers beneath the nanotube as well as diffusion of carriers  when the laser illuminates away from the nanotube.
 Upper inset:  Photocurrent as a function of position perpendicular to the long axis of nanotube. Photocurrent is measured up
 to 1~mm away from the nanotube.
 Lower inset:  SEM image of the $\sim500~\mu$m nanotube device, where the dotted line indicates the position of the
 nanotube. The arrow shows the position and direction of the scanned laser.}
 \label{PCvPosn}
\end{figure}

It is interesting to note that the photo-voltage depends only weakly
on where the laser illuminates the sample relative to the nanotube.
This fact is illustrated in the upper inset to Fig.~\ref{PCvPosn},
which shows the photocurrent as a function of position perpendicular
to the long axis of the nanotube at the position indicated by the
arrow in the lower inset.  The curve is slightly asymmetric about
the position of the nanotube due to the proximity of the device to
the edge of the silicon chip.
The surprising feature of Fig.~\ref{PCvPosn} is that there is an
appreciable photocurrent, reduced only by a factor of $\sim3$, even
when the laser spot is centered over 1~mm away from the device.
Using a laser spot with an $\sim60~\mu$m gaussian width, there are
virtually no carriers produced near the nanotube at such a large
distance from the nanotube.  Since the laser isn't producing
carriers near the nanotube, the carriers must be mobile in the Si.
Thus the photocurrent that is induced by the photovoltage arises
from laser generated carriers near the nanotube as well as carriers
that diffuse towards the nanotube.  We can phenomenologically
express the number of carriers under the nanotube as
\begin{equation}
 \Delta n(x)/\Delta n_o=Ae^{-2x^2/\omega^2}+Be^{-|x|/L_n},
 \label{eqnGenerationMobility}
\end{equation}
where the first term accounts for direct photo-generation of
carriers in the silicon under the nanotube, and the second term is
from carriers generated at some distance from the nanotube diffusing
towards the nanotube. Using Fig.~\ref{PCvsIntensity} to relate the
photocurrent to incident power, and since $\Delta n$ is proportional
to the laser power, we can relate the measured photocurrent to the
ratio of $\Delta n(x)/\Delta n_o$, where $\Delta n_o$ is the number
of carriers generated in the silicon at the photocurrent maxima. The
solid line in Fig.~\ref{PCvPosn} is a fit to the experimental data
using equation Eq.~\ref{eqnGenerationMobility}, with the exponential
prefactors $A$ and $B$, the beam waist $\omega$, and the diffusion
length $L_n$ as free parameters.  The experimental data is well fit
by Eq.~\ref{eqnGenerationMobility} using $\omega=58~\mu$m and
$L_n=350~\mu$m.  The extracted value of $L_n$ is an effective
diffusion length, as opposed to the bulk diffusion length, and can
depend greatly on the detailed geometry and injection conditions
that are present.  These results illustrate that photo-gating is a
long-range interaction.  Carriers that are generated in the silicon
substrate millimeters away from the nanotube still have a
significant effect on the nanotube circuit.

Photo-gated nanotube devices present an interesting alternative to
photo-detection on the nanoscale.  Such a device would separate the
absorption of photons from the detection current, such that each can
be independently optimized.  The gate material (silicon in this
case) would determine the absorption properties of the detector, and
could be modified to increase the absorption cross-section at the
desired wavelengths.  The sensitivity of the photo-gated nanotube
detector is determined by the nanotube-gate coupling $\frac{\partial
I_{SD}}{\partial V_g}$, which may be custom tailored through device
engineering of the Schottky barriers,\cite{HeinzeAPL} and geometry,
as well as mode of operation (sub-threshold vs.\
on).\cite{AppenzellerPRL} Nanotubes offer many properties that make
them superior channel materials such as high
mobilities\cite{DurkopNanoLett} and large saturation
velocities,\cite{PenningtonPRB} but there is no reason to expect
that photo-gating of the type studied here should be unique to
carbon nanotube channels.

%summary begins here
In conclusion we have demonstrated that light absorption in the
substrate of carbon nanotube field effect transistors generates an
extra photo-voltage at the Si/SiO$_2$ interface.  This photovoltage
acts as an additional negative gate voltage of order 10-100~mV on
the nanotube transistor and, depending on the magnitude of the
nanotube-gate coupling, produces changes in the source-drain current
of up to $\sim2$~nA which is considerably larger than previously
reported photocurrents in carbon
nanotubes.\cite{FreitagNanoLett,OhnoAPL} The photo-voltage responds
nonlinearly to the power in the laser beam, increasing only by a
factor of 3 over three orders of magnitude in laser power, and is
only weakly dependent on where the carriers are generated with
respect to the nanotube. We find that charge generated in the Si
substrate up to 1~mm away from the nanotube still produces a
measurable photocurrent. The current due to photogating can, in many
cases, be the dominant component of the photocurrent measured
through nanotube transistors and must be accounted for in order to
determine the intrinsic/direct photocurrent in nanotube devices.
Though photogating complicates the analysis of photocurrent
measurements, it also offers a different mode of photodetection
where the absorption and detection elements are separate.  This
style of detection would allow for individual optimization of the
photoabsorber and the electronic sensitivity of the transistor
channel.

\begin{acknowledgments}
We would like to acknowledge Jennifer Sebby-Strabley for useful
discussions.  We would also like to acknowledge funding from the NSF
CAREER program under grant number \mbox{DMR-0094063}, the NSF MRSEC
program under grant number \mbox{DMR-0520527}, and the NSF NSEC
program under grant number \mbox{DMR-0425880}.
\end{acknowledgments}

\appendix*
\section{Dember Voltage}
In addition to a potential forming from the redistribution of
photo-excited charge, another voltage that can form in the
substrate, the Dember voltage, is particularly relevant for high
injection conditions where $\Delta n\gtrsim p_o$. The Dember voltage
is produced when photo-excited electrons and holes diffuse at
different rates.  The differing diffusion rates lead to a separation
of the charge, which in turn generates an electric field in the
silicon substrate.  Because the electron mobility is higher than the
hole mobility in silicon, the sign of the Dember voltage is always
positive and will appear in series with the band-induced
photo-voltage, effectively reducing the total amount of photovoltage
measured in the nanotube device. The Dember voltage is expressed
as\cite{KronikSurfSciRep}
\begin{equation}
 V_d=\frac{k_BT}{e}\frac{b-1}{b+1}\ln[1+\frac{(b+1)\Delta
 n}{n_o+bp_o}]
 \label{eqnDember}
\end{equation}
where $b$ is the ratio of electron to hole mobility
$\mu_n/\mu_p\sim1.8$ and $n_o$ is the equilibrium number of
electrons in the silicon. Using the over-estimated value of $\Delta
n$ from Eq.~\ref{eqnLowInjection}, the Dember voltage in our devices
is $V_d\approx1$~mV which is much less than the measured
photo-voltage due to band curvature for the analysis presented here.
For larger values of excited carriers, the Dember voltage can become
comparable to the photo-voltage and should not be ignored. The
effect of the Dember voltage is shown in Fig.~\ref{PVdN}, where the
calculated total photovoltage for the case when there is only band
curvature (solid line) is compared to when the Dember voltage is
included (dotted line), assuming a bulk doping density of
$N_a\approx1.2\times10^{18}~\text{cm}^{-3}$ and initial band bending
of $\phi_{so}=-120$~mV. As can be seen, the effect of the Dember
voltage is small under low-injection conditions, but becomes sizable
when the number of injected carriers nears the bulk doping density.
 It is important to note that both the band curvature and Dember voltages depend on the bulk
doping and oxide charge densities and must be recalculated if higher
or lower doping concentrations are used.

\newpage
\textbf{List of Figures:}
\begin{enumerate}
 \item Source-drain ($I_{SD}$) as a function of back gate voltage.  The conduction through the nanotube increases at
 large values of both negative and positive gate voltages.  Inset:  AFM image of a $\sim5~\mu$m long single wall carbon nanotube
 transistor device.\\
 \item Photocurrent measured through the nanotube channel as a function of gate voltage.
The photocurrent switches sign near $V_g=0$.
Inset:  Schematic of the energy bands with initial band bending at the Si/SiO$_2$ interface.\\
 \item Schematic band diagrams showing band bending $\phi_s$ in the silicon substrate for the as fabricated device (a) and under
laser illumination (b).  For typical oxide charge densities, the
initial band bending $\phi_{so}$ is $\sim100-200$~mV. Under
illumination, the positive oxide charge is partially screened by
photogenerated electrons that become trapped at the interface,
reducing the surface potential from
$\phi_{so}\rightarrow\phi_{s\Lambda}$.  The resulting photovoltage
$\delta V_{PV}$ is given by Eq.~\ref{eqnVPV}.\\
 \item Calculated magnitude of photovoltage as a function of the number of photo-excited carriers, assuming
a bulk doping density of $N_a\approx1.2\times10^{18}~\text{cm}^{-3}$ and an initial surface potential of
$\phi_{so}\sim120$~mV.  The solid line includes only the effect of the initial band bending.  The dotted line
includes the effect of the Dember voltage (see appendix).
Inset:  Calculated normalized electric field at the Si/SiO$_2$ interface as a function of the surface potential
and injection ratio.  The dotted line indicates the value of the field $F_o$ as determined from the initial charge
in the silicon $Q_s$\\
 \item Numerically calculated derivative from the data in Fig.~\ref{IVg}.  The form of the derivative is extremely similar
to the measured photocurrent in Fig.~\ref{PCVg} where the p-channel conduction is increased and the n-channel is suppressed.\\
 \item Measured photocurrent as a function of the incident laser power on a long channel nanotube device.
The non-linearity of the intensity dependence indicates that the photocurrent does not arise from a direct process.\\
 \item Ratio of photo-generated electrons in the silicon substrate as a function of position.  The circles are the
experimentally measured values, extracted by relating the raw photocurrent vs. position in the upper inset to the intensity dependence
shown in Fig.~\ref{PCvsIntensity}.  The line is a fit to Eq.~\ref{eqnGenerationMobility} which includes direct generation
of carriers beneath the nanotube as well as diffusion of carriers  when the laser illuminates away from the nanotube.
Upper inset:  Photocurrent as a function of position perpendicular to the long axis of nanotube. Photocurrent is measured up
to 1~mm away from the nanotube.
Lower inset:  SEM image of the $\sim500~\mu$m nanotube device, where the dotted line indicates the position of the
nanotube. The arrow shows the position and direction of the scanned laser
\end{enumerate}


\begin{thebibliography}{37}
\bibitem[{Ban()}]{BandGaps}
\bibinfo{note}{Nanotubes with diameters larger than $\sim0.5~nm$ have band gaps
  that are less than the photon energy of our laser.}
\bibitem[{Iro()}]{IronNitrateGrowth}
\bibinfo{note}{This device was grown using an iron nitrate catalyst at
  900$^\text{o}$C, with 500~sccm hydrogen, 1000~sccm methane, and 20~sccm
  ethylene.}
\end{thebibliography}
\end{document}